\documentclass[12pt, openright, oneside, a4paper, english]{abntex2}

\usepackage{lmodern}			
\usepackage[T1]{fontenc}		
\usepackage[utf8]{inputenc}		
\usepackage{lastpage}			
\usepackage{indentfirst}		
\usepackage{color}				
\usepackage{graphicx}			
\usepackage{microtype} 			
\usepackage{amsthm}
\usepackage{amsmath}
\usepackage{mathtools}
\usepackage{amsfonts}
\usepackage{amssymb}
\usepackage{caption}
\usepackage{subcaption}
\usepackage{tabularx}
\usepackage{makecell}
 \usepackage[frozencache=true,cachedir=minted-cache]{minted} 
\renewcommand{\listoflistingscaption}{List of Program Code}

\usepackage{amsmath}

\usepackage[ruled,lined,linesnumbered]{algorithm2e}

\usepackage{multirow}

\usepackage[brazil]{babel}
\addto\captionsbrazil{%

\renewcommand{\contentsname}{Contents}

}

\definecolor{LightGray}{gray}{0.9}
\usepackage{lipsum}				

\usepackage[alf]{abntex2cite}	

\titulo{Bridging Swift Error Handling Model to C++}
\autor{Roberto Gomes Rosmaninho Neto}
\local{Belo Horizonte, Minas Gerais}
\data{2022}
\orientador{Fernando Magno Quintão Pereira}

\instituicao{%
 Universidade Federal de Minas Gerais 
 \par Instituto de Ciências Exatas
 \par Departamento de Ciência da Computação}
\tipotrabalho{Doctorate Thesis}
\preambulo{Final Computing Project Report to the Bachelor Degree in Computer Science of the Universidade Federal de Minas Gerais}

\definecolor{blue}{RGB}{41,5,195}

\makeatletter
\hypersetup{
		pdftitle={\@title}, 
		pdfauthor={\@author},
    	pdfsubject={\imprimirpreambulo},
	    pdfcreator={LaTeX with abnTeX2},
		pdfkeywords={abnt}{latex}{abntex}{abntex2}{trabalho acadêmico}, 
		colorlinks=true,       		
    	linkcolor=blue,          	
    	citecolor=blue,        		
    	filecolor=magenta,      		
		urlcolor=blue,
		bookmarksdepth=4
}
\makeatother

\makeindex

\begin{document}

\frenchspacing 

\pretextual

\imprimircapa
\cleardoublepage

\imprimirfolhaderosto



\newpage
\begin{center}
{\ABNTEXchapterfont\Large\textsc{Statement of Authorship}}
\end{center}


I hereby declare that the thesis submitted is my own work. All direct or indirect
sources used are acknowledged as references. I further declare that I have not submitted this thesis at any other institution in order to obtain a degree.


\newpage
\begin{dedicatoria}
  \vspace*{\fill}
  \centering
  \noindent
  \textit{ To my friends and family, the reason of everything.} \vspace*{\fill}
\end{dedicatoria}

\begin{agradecimentos}

I would like to thank my advisor, Fernando Pereira, who provided me with the best opportunities in my life. My colleagues, Angelica Moreira and Breno Campos, who have always supported me on this path, and my parents, Roberto and Rogeria Rosmaninho, for helping me get this far!

\end{agradecimentos}

	
		


\setlength{\absparsep}{18pt} 
\begin{resumo}[Abstract]
 \begin{otherlanguage*}{english}

Error handling is the process of responding to and recovering from error conditions in the program. In Swift, errors are represented by values of types that conform to the Error protocol. Throwing an error lets you indicate that something unexpected happened, and the normal flow of execution can’t continue. A throw statement is used to throw an error. Optional returns are used to represent the absence of a value, but when an operation fails, it’s often useful to understand what caused the failure so that code can respond accordingly. Therefore, I propose to bridge the Swift Error Handling modeling to C++ to improve the interoperability between the programming languages. The main idea is to be able to throw a C++ exception that stores a thrown Swift Error that has to be represented by a C++ class. In addition to that, to support C++ programs that don't use exceptions, I propose an additional interoperability mode for throwing functions. When C++ exceptions are disabled, C++ functions should return a result value that contains either the value returned by the function, or the Swift Error value: the Swift::Exception<T> class.


\end{otherlanguage*}
\end{resumo}

\newpage


\listoflistings

\pdfbookmark[0]{\contentsname}{toc}
\tableofcontents*
\cleardoublepage

\textual

\chapter{Introduction}

Swift is a general-purpose programming language built using a modern approach to safety, performance, and software design patterns. On December 3, 2015, the Swift language, supporting libraries, debugger, and package manager were published under the Apache 2.0 license with a Runtime Library Exception. The source code is hosted on GitHub where it is easy for anyone to get the code, build it themselves, and even create pull requests to contribute code back to the project. \cite{swiftorg} The effort to improve and optimize the Swift Compiler has been one of the main goals of the Swift community, mainly discussing ideas, goals, implementations, and implementation on specific topics at the Swift Online Fórum.

The growing number of APIs and projects using the Swift Programming Language has been attracting attention not only from the Swift community but also from the C++ community as well. Therefore, if anyone from the C++ community wants to use some API developed in Swift or if they want to incrementally change the codebase of their application from C++ to Swift, they should have the ability to do that by relying on interoperability between the two languages. That's the reason that makes C++ interoperability with Swift so important.

In order to use Swift APIs or incrementally adopt Swift in a C++ code, developers must ensure that their application will not crash due the errors from C++ code or errors from Swift code. Handling errors from C++ in C++ code is straightforward. On the other hand, if a Swift function used in a C++ codebase through a bridging header throws an Error, it's currently impossible to know what happened from the C++ side. So, the program cannot handle errors from Swift and may crash without any meaningful message.


This project aimed to develop and implement the infrastructure on the Swift Compiler to extend the Swift handling errors to C++ through a bridging header.

\section{General Goals}
This project aimed to extend the Swift compiler to allow Swift functions to propagate errors to C++ code. The first modification was to modify the current implementation to enable casting the ``Swift::Error'' C++  Class to the user’s defined errors. The second modification provided another C++ class that also wraps the Swift Error but returns it wrapped in a \textit{std::expected}\cite{cppException} like implementation to be used by users that disable the C++ exceptions to handle the error in their way instead of throwing it.

\section{Specific Goals}
The goals of this projects can me summarized in:
\begin{itemize}
    \item Create a new C++ class, like \textit{std::exception}, and stores a Swift error value.
    \item Extend this C++ Swift Error representation to cast to the user's defined errors.
    \item Generate a C++ class similar to \textit{std::expected}\cite{cppException} to provide error handling for clients that don't use C++ exceptions.
    \item Automatically switch the error handling model regarding the cpp\_exceptions flag value.

\end{itemize}

\chapter{Theoretical References}
\section{History}
The history of the C++ programming language starts in 1979, when Bjarne Stroustrup was doing work for his Ph.D. thesis. He began working on "C with Classes", which as the name implies was meant to be a superset of the C language. His goal was to add object-oriented programming into the C language. His language included classes, basic inheritance, inlining, default function arguments, and strong type checking in addition to all the features of the C language. \cite{historyCpp}

The first C with Classes compiler was called Cfront, which was derived from a C compiler called CPre. It was a program designed to translate C with Classes code to ordinary C. Cfront would later be abandoned in 1993 after it became difficult to integrate new features into it, namely C++ exceptions. \cite{historyCpp}

\section{Exceptions}
Exceptions provide a way to react to exceptional circumstances (like runtime errors) in programs by transferring control to special functions called handlers.

To catch exceptions, a portion of the code is placed under exception inspection. This is done by enclosing that portion of code in a try-block. When an exceptional circumstance arises within that block, an exception is thrown that transfers the control to the exception handler. If no exception is thrown, the code continues normally, and all handlers are ignored. \cite{cppException}

An exception is thrown by using the throw keyword from inside the \textbf{\textit{try}} block. Exception handlers are declared with the keyword \textbf{\textit{catch}}, which must be placed immediately after the try block:

\begin{listing}[H]
\begin{minted}[fontsize=\small,frame=single,framesep=2mm,
                baselinestretch=1.2,fontsize=\footnotesize,
                linenos]{C++}
// exceptions
#include <iostream>
using namespace std;

int main () {
  try {
    throw 20;
  } catch (int e) {
    cout << "An exception occurred. Exception Nr. " << e << '\n';
  }
  return 0;
}
\end{minted}
\caption{Example of the Exception Handling in C++.}
\label{lst:example3}
\end{listing}

The code under exception handling is enclosed in a try block. In this example, this code simply throws an integer number as an exception:

\begin{listing}[H]
\begin{minted}[fontsize=\small,frame=single,framesep=2mm,
                baselinestretch=1.2,fontsize=\footnotesize,
                linenos]{C++}
throw 20;
\end{minted}
\caption{Example of a simple throw statement.}
\label{lst:example4}
\end{listing}

\section{Swift Error Handling Model}
In Swift, errors are represented by values of types that conform to the Error protocol. Throwing an error indicates that something unexpected happened, and the normal flow of execution can’t continue. 

When a function encounters an error condition, it throws an error. That function’s caller can then catch the error and respond appropriately. \cite{swiftExceptions}

\begin{listing}[H]
\begin{minted}[fontsize=\small,frame=single,framesep=2mm,
                baselinestretch=1.2,fontsize=\footnotesize,
                linenos]{Swift}
enum numError: Error {
    case twenty
}

func main () throws -> Int {
    do {
        throw numError.twenty
    } catch {
        print("An exception occurred. Exception Nr. \(error)");
    }
    return 0;
}
\end{minted}
\caption{Example of declaration of a function that can throw an error. This is almost an equivalent program to \ref{lst:example3}.}
\label{lst:example5}
\end{listing}

A function indicates that it can throw an error by including the \textbf{\textit{throws}} keyword in its declaration. The keyword \textbf{\textit{throw}} was used in C++ as well, at least until the C++17 standard.\cite{throwCpp} However, unlike Swift, in C++, this keyword works as a dynamic exception specification followed by a type indicating that the function with it may throw exceptions of that type or a type derived from it. On the other hand, if the function is followed by \textbf{\textit{throw()}}, the compiler knows that it can't throw an exception.

When you call a function that can throw an error, you prepend the try keyword to the expression. In Swift, this structure is called \textbf{\textit{do-try-catch}}. This structure is very common in codebases that choose to work with exceptions.

\begin{listing}[H]
\begin{minted}[fontsize=\small,frame=single,framesep=2mm,
                baselinestretch=1.2,fontsize=\footnotesize,
                linenos]{C++}
do {
    try canThrowAnError()
    // no error was thrown
} catch {
    // an error was thrown
}
\end{minted}
\caption{Example of the correct implementation to call a function that can throw an error.}
\label{lst:example6}
\end{listing}

If a function throws an error, then Swift automatically propagates this error out of its current scope until a catch clause handles it, even if this function is called from a C++ program. That is one of the reasons that motivate this work to improve the interoperation between Swift and C++ by creating two solutions to handle Swift Errors in C++.

\section{The std::expected<T,E>}

The proposal p0323r3\cite{p0323r3} presents the class template \textit{\textbf{expected<T, E>}} that contains either:
\begin{itemize}
    \item A value of type T, the expected value type; or
    \item A value of type E, an error type used when an unexpected outcome occurred.
\end{itemize}

The interface can be seen as whether the underlying value is the expected value (of type T ) or an unexpected value (of type E ). The original idea comes from Andrei Alexandrescu C++ and Beyond 2012: Systematic Error Handling in C++ Alexandrescu\cite{AndreiExpected}. The interface and the rationale were based on \textit{\textbf{std::optional}} \cite{N3793}. The proposal's author considers \textit{\textbf{expected<T, E>}} as a supplement to \textit{\textbf{optional<T>}}, expressing why an expected value isn't contained in the object.

It's important to recap that C++'s two main error mechanisms are exceptions and return codes. Characteristics of a good error mechanism are\cite{p0323r3}:

\begin{enumerate}
    \item Error visibility: Failure cases should appear throughout the code review: debugging can be painful if errors are hidden.
    \item Information on errors: Errors should carry information from their origin, causes, and possibly the ways to resolve them.
    \item Clean code: Treatment of errors should be in a separate layer of code and as invisible as possible. The reader could notice the presence of exceptional cases without needing to stop reading.
    \item Non-Intrusive error Errors should not monopolize the communication channel dedicated to normal code flow. They must be as discrete as possible. For instance, the return of a function is a channel that should not be exclusively reserved for errors.
\end{enumerate}

The same analysis can be done for the  \textit{\textbf{expected<T, E>}} class and observe the advantages over the classic error reporting systems.\cite{p0323r3}
\begin{enumerate}
    \item Error visibility: It takes the best of the exception and error code. It's visible because the return type is  \textit{\textbf{expected<T, E>}}, and users cannot ignore the error case if they want to retrieve the contained value.
    \item Information: Arbitrarily rich.
    \item Clean code: The monadic interface of expected provides a framework delegating the error handling to another layer of code. Note that  \textit{\textbf{expected<T, E>}} can also act as a bridge between an exception-oriented code and a nothrow world.
    \item Non-Intrusive Use the return channel without monopolizing it.
\end{enumerate}

\chapter{Methodology}


\section{Current State of the C++ and Swift Interoperability}
The interoperability between Swift and C++ programming languages works through a bridging header. This header is a C++ file containing all the function signatures, types, and class declarations that a user wants to export from a Swift source code to a C++ program. To generate this header, we must modify the Swift Compiler to correctly recognize and represent the information we want to export using C++ types, macros, functions, and other useful resources.

\section{Initial implementation}
The main goal of this work is to continue translating the Swift Error Handling to C++ Error Handling.

To achieve this goal, I got in touch with Alex Lorenz, the Apple employee assigned to work with students in the Swift Interoperability Work-group\cite{cppInteropWorkgroup}. The group is composed of community members and Apple employees from the Clang team. We did sync-up meetings every week to ask questions about any blockers we had on our projects, to know each one and the projects better, and to get guidelines from Apple's perspective. Also, the Work-group has a Slack channel public where I continued in touch with Alex to ask simple implementation questions and to discuss about GitHub CI tests and reviews.

The first step was to study Swift and C++ handling error model individually from the user’s perspective, write the example \ref{lst:example3} using both programming languages and analyze the syntax of each one, as shown in the Theoretical References section. Then, I started studying the Swift Compiler, especially the Code Generation part on the PrintAsClang\cite{printAsClang} directory.

The second step, and my first official contribution\cite{PR1}, following a Test-Driven Development, was to create tests with functions that throw an error. The idea was to implement a simple Enum Error with two cases and two functions: the first marked with ``throws'' flag but never throwing any error, and the second with ``throws'' as well, but throwing an error from the Enum error created earlier.

The third step, and my second official contribution\cite{PR2}, was to modify the compiler to delete all flags that classify every function as a ``noexcept'' function. In the previous implementation and from a C++ point of view, no function could throw errors. The solution to solve this hardcoded behavior was to verify during the header code generation if a function was declared in Swift with the flag ``throws'' or not. This modification made the tests on \cite{PR1} pass as expected, as they were functions that could throw in Swift, but were marked as ``noexcept'' in the C++ bridge header.

Unfortunately, this last step was not straightforward as expected. The ``GitHub CI'' pointed out that one single test from the entire Test Suit failed on the architecture ``i386-apple-watchos2.0-simulator''. After some debugging, we realized that two default parameters were missing on the function signature, the first called “self” and the second called ``error''. The former was crucial to the next steps as it contains the information of an error if one was thrown during the execution of the function.

\begin{listing}[H]
\begin{minted}[fontsize=\small,frame=single,framesep=2mm,
                baselinestretch=1.2,fontsize=\footnotesize,
                linenos]{C++}
inline void throwFunction() {
  void* opaqueError = nullptr;
  void* self = nullptr;
  _impl::$s9Functions13throwFunctionyyKF(self, &opaqueError);
  if (opaqueError != nullptr)
    throw (swift::_impl::NaiveException("Exception"));
}
\end{minted}
\caption{Example of a function that can throw a NaiveException if the Swift function correspondent throws an error.}
\label{lst:exampleThrowFunction}
\end{listing}

The above example illustrates the third contribution\cite{PR3} of this project:
\begin{itemize}
    \item The ``if condition" testing if the ``opaqueError'' was modified during the execution of the Swift function. 
    \item The NaiveException C++ class. 
\end{itemize}

The first modification tests if the Swift function throws an error. In this case, the ``opaqueError'' parameter holds the information needed to identify the thrown error instead of ``nullptr''. The second modification implements a C++ class to construct a simple object with a message and throw it to the C++ caller function. Also, this message can be accessed with the ``\textbf{\textit{getMessage()}}'' method.

\newpage

\section{Final Implementation}
The first part of this project aimed to provide a minimal version of the Swift handling error model to C++, so a C++ function caller could know that a Swift function called throws an error. The second and final part of this project aimed to specialize that knowledge by providing a general ``\textbf{\textit{Swift::Error}}'' representation, an infrastructure to recognize which error case the function threw, and a new C++ class \textbf{\textit{Swift::Expected}}'' to return the error thrown. 

The fourth contribution\cite{PR4} of this project was the mentioned general ``\textbf{\textit{Swift::Error}}'' representation to replace the ``NaiveException'' on the ``throw'' statement of the C++ function representation in the generated bridging header. This class implements its constructor and destructor using the functions ``\textbf{\textit{swift\_errorRetain}}'' and ``\textbf{\textit{swift\_errorRelease}}'' to handle the Swift Error properly avoiding unexpected behaviors and memory leaks. The main idea of this class was to provide a representation for the Swift error type that can be examined on the C++ side. This representation behaves as Swift protocol type in C++, and is able to represent all Swift error values in C++.

The main idea of this class was to replace the need for using the ``std::exception'' library, that currently has dependencies that can't be imported into the generated bridging header, and to work as a ``superclass'' for all Swift Error in C++.

\begin{listing}[H]
\begin{minted}[fontsize=\small,frame=single,framesep=2mm,
                baselinestretch=1.2,fontsize=\footnotesize,
                linenos]{Swift}
class TestDestroyed {
  deinit {
    print("Test destroyed")
  }
}

public struct DestroyedError : Error {
  let t = TestDestroyed()
}

public func testCast(_ e: Error) {
    let _ = e as? DestroyedError
}
\end{minted}
\caption{Swift function that can throw an Error, but doesn't have a throw statement.}
\label{lst:example7}
\end{listing}

The fifth contribution\cite{PR5} was the most challenging part of this project. It required debugging the example \ref{lst:example7} at the assembly level using the lldb\cite{lldb} trace and breakpoints to discover the information needed to reproduce the casting at line 12 using our C++ infrastructure. 

In a 1-1 debugging meeting with Alex from Apple, we found out that to dynamically cast an error we need some metadata information, a symbolic type, and external global variables. Some of these required pieces of information could be hardcoded, and others we could get by declaring them as ``external "C" '' functions and variables since the Swift Compiler uses them to build the LLVM IR.

Finally, after implementing all the auxiliary functions to dynamically cast the ``\textbf{\textit{Swift::Error}}'' to a concrete Swift type that represents the case thrown, I implemented the ``\textbf{\textit{as<T>()}}'' function as a method from the ``\textbf{\textit{Swift::Error}}'' class to provide the user a natural experience to handle errors in C++.

In the sixth contribution \cite{PR6}, I improved the mentioned dynamic cast to return a ``\textbf{\textit{Swift::Optional<T>}}'' type within a value if the casting was successful or a ``\textbf{\textit{Swift::Optional<T>::none()}}'' if not. Also, I deleted the ``\textbf{\textit{NaiveException}}'' that wasn't necessary anymore and changed ``\textbf{\textit{Swift:Error}}'' implementation to the same file where ``\textbf{\textit{Swift::Optional<T>}}'' was defined to be able to modify its return type.

Finally, in the seventh and last contribution\cite{PR7}, I developed and implemented my own version of the ``\textbf{\textit{std::Expected<T>}}''\cite{p0323r4} proposal as ``\textbf{\textit{Swift::Expected<T>}}'' and implemented the support for the bridging header chose the proper way to handle a Swift thrown error according to the C++ user preference to use or not exceptions by passing the ``\textit{-fno-exceptions}'' flag to the compiler. The first part of this project was very challenging due to the responsibility to build a great API that will be used extensively by the community. Also, this is the implementation of a proposal that has been discussed for years in the community, with only a few people risking implementing their version of ``\textbf{\textit{std::Expected<T>}}''. 

Also, two important requirements from Apple were to use the ``\textbf{\textit{Swift:Error}}'' as the only type of error that could be returned and to store at the same buffer the error or the value of type T to optimize the memory for objects from this class. The last part of this contribution wasn't trivial either: in order to avoid code repetition either on the compiler and on the bridging header, I used an alias (\textit{ThrowingResult}) and a Macro \textit{(SWIFT\_RETURN\_THUNK}) to specify which implementation of the part of the C++ implementation would be used to properly represent the Swift function regarding the error modeling choose by the C++ user as shown below:

\begin{listing}[H]
\begin{minted}[fontsize=\small,frame=single,framesep=2mm,
                baselinestretch=1.2,fontsize=\footnotesize,
                linenos]{C++}
#ifdef __cpp_exceptions
 template<class T>
 using ThrowingResult = T;
 #define SWIFT_RETURN_THUNK(T, v) v
 #else
 template<class T>
 using ThrowingResult = Swift::Expected<T>;
 #define SWIFT_RETURN_THUNK(T, v) Swift::Expected<T>(v)
 #endif
\end{minted}
\caption{Macro definition to replace the function type and return regarding the use of exceptions.}
\label{lst:ThrowingResult}
\end{listing}




\chapter{Results and Example}
The contributions cited above show that we successfully accomplished our primary goals. Finally, using two approaches, C++ users can handle errors thrown by Swift functions. Below, the straightforward example shows a Swift function that can throw an error to avoid unexpected behavior.

\begin{listing}[H]
\begin{minted}[fontsize=\small,frame=single,framesep=2mm,
                baselinestretch=1.2,fontsize=\footnotesize,
                linenos]{Swift}
@_expose(Cxx)
public enum DivByZero : Error {
    case divisorIsZero
    case bothAreZero

    // Function to print the case thrown
    public func getMessage() {
        print(self)
    }
}

@_expose(Cxx)
public func division(_ a: Int, _ b: Int) throws -> Float {
    if a == 0 && b == 0 {
        throw DivByZero.bothAreZero
    } else if b == 0 {
        throw DivByZero.divisorIsZero
    } else {
        return Float(a / b)
    }
}
\end{minted}
\caption{Example of a user-defined Error and a Swift function that can throw it. }
\label{lst:exampleSwift}
\end{listing}

The client of this function, the C++ user, must first compile the program in which this function is defined to create the bridging header to export the Swift implementation to C++. The below command can be used to generate this header; it was extracted from the Swift test suit and simplified as much as possible. 

\begin{listing}[H]
\begin{minted}[fontsize=\small,frame=single,framesep=2mm,
                baselinestretch=1.2,fontsize=\footnotesize,
                linenos]{bash}
$ swift-frontend Example.swift -typecheck -module-name Functions \
    -enable-experimental-cxx-interop -emit-clang-header-path functions.h
\end{minted}
\caption{Command to create the bridging header of a Swift program.}
\label{lst:exampleCommandHeader}
\end{listing}

This command takes the ``Example.swift'' file as input and outputs the ``functions.h'' file with the C++ representation of the Swift functions under the ``Functions'' namespace.

\begin{listing}[H]
\begin{minted}[fontsize=\small,frame=single,framesep=2mm,
                baselinestretch=1.2,fontsize=\footnotesize,
                linenos]{C++}
inline Swift::ThrowingResult<float> division(swift::Int a, swift::Int b) {
  void* opaqueError = nullptr;
  void* _ctx = nullptr;
  auto returnValue = _impl::$s9Functions8divisionySfSi_SitKF(a,b,_ctx,&opaqueError);
  if (opaqueError != nullptr)
#ifdef __cpp_exceptions
    throw (Swift::Error(opaqueError));
#else
    return SWIFT_RETURN_THUNK(float, Swift::Error(opaqueError));
#endif

  return SWIFT_RETURN_THUNK(float, returnValue);
}
\end{minted}
\caption{Example of using a Swift function that can throw an Error in C++ that can be caught.}
\label{lst:exampleHeader}
\end{listing}

 The C++ function represented  generated by the compiler, and shown above, can be explained in four parts:

\begin{itemize}
    \item \textbf{Function signature}: The function has the type ``\textbf{\textit{ThrowingResult<float>}}'' instead of the ``\textbf{\textit{float}}'' type declared on Swift. The implementation of this type is shown on listing \ref{lst:ThrowingResult}, and it is used to avoid code repeating by declaring the function twice with different types to satisfy the two error handling approaches. 
    \item \textbf{The Swift function call}: This has two extra arguments we defined before; the first holds the context, and the second the error information. They are required to match the Swift ABI, and the last argument has the core information to specify if the Swift function threw an error and which error it is or if it returned as expected.
    \item \textbf{The opaqueError is nullptr?}: This if condition checks if the ``\textbf{\textit{opaqueError}}'' argument was modified during the execution of the function to hold Swift error information or not. Also, inside the branch, we have two different implementations regarding the use of exceptions by the C++ user; if it's enabled, we throw a ``\textbf{\textit{Swift::Error}}'' object initiated with the ``\textbf{\textit{opaqueError}}'' argument. If not, we \textbf{return} a macro that will be replaced by: \\  ``\textbf{\textit{Swift::Expected<float>(Swift::Error(opaqueError))}}''.
    \item \textbf{The return expression}: The macro returned here will be replaced at compile time depending on the use of C++ exceptions. It can either only returns the ``\textbf{\textit{returnValue}}'' or the ``\textbf{\textit{Swift::Expected<float>(returnValue)}}'' that contains the value we expected instead of an Error as it previous use in this example.
\end{itemize}

The C++ programmer uses this function representation by importing the generated Swift bridging header to its program. Below we show two uses of this function using C++ exceptions:

\begin{listing}[H]
\begin{minted}[fontsize=\small,frame=single,framesep=2mm,
                baselinestretch=1.2,fontsize=\footnotesize,
                linenos]{C++}
#include <cassert>
#include <cstdio>
#include "functions.h"

int main() {

    // This example catches an exception
    try {
        auto result = Functions::division(0,0);
        printf("result = %f\n", result);
    } catch (Swift::Error& e) {
        auto errorOpt = e.as<Functions::DivByZero>();
        assert(errorOpt.isSome());

        auto errorVal = errorOpt.get();
        assert(errorVal == Functions::DivByZero::bothAreZero);
        errorVal.getMessage();
    }

    // This example gets the correct value returned by the function 
    try {
        float result = Functions::division(4,2);
        printf("result = %f\n", result);
    } catch (Swift::Error& e) {
        auto errorOpt = e.as<Functions::DivByZero>();
        assert(errorOpt.isSome());

        auto errorVal = errorOpt.get();
        errorVal.getMessage();
    }
    return 0;
}
\end{minted}
\caption{Examples of using a Swift function that can throw an error in C++ and handles it.}
\label{lst:exampleCppExceptions}
\end{listing}

In this example, we can see how to catch the ``\textbf{\textit{Swift::Error}}'', dynamically cast it to the user-defined error, which returns an Optional type that says if the casting was successful or not, check if it threw the case that we expected, and then, tells the user name of it using the ``\textbf{\textit{getMessage()}}'' function. If the function doesn't throw an error, we get the ``\textbf{\textit{float}}'' result and print it to the standard output.

The commands below show how to compile and test this program. First, we need to compile and generate its object file, compile it with the Swift file containing the function and error definitions, and then generate the executable program.

\begin{listing}[H]
\begin{minted}[fontsize=\small,frame=single,framesep=2mm,
                baselinestretch=1.2,fontsize=\footnotesize,
                linenos]{Bash}
$ clang++ -I ${SWIFT_BUILD_DIR}/swift-macosx-arm64/./lib/swift \ 
        -c Example.cpp -I /. -o example.o

$ swiftc -Xfrontend -enable-experimental-cxx-interop Example.swift -o example \
        -Xlinker example.o -module-name Functions -Xfrontend \
        -entry-point-function-name -Xfrontend swiftMain

$ ./example

\end{minted}
\caption{Commands to build, link and execute a C++ program that calls a Swift function and can throw exceptions }
\label{lst:exampleCommandExceptions}
\end{listing}

The second approach to handle Swift errors in C++ was inspired by the C++ proposal ``p0323r4 std::expected''\cite{p0323r4}. This approach required a new C++ parameterized class called ``Swift::Expected<T>'', which can be initialized to hold either a Swift::Error or a value of type T, never simultaneously. This class has auxiliary functions to improve the interface with the user and simplify the test and get an error or a value. An instance of this class is returned by a C++ representation of a Swift function that can throw an error, but the C++ user disabled exceptions at compile time.

The example below uses the same Swift function and C++ representation of it but with this new error-handling model:


\begin{listing}[H]
\begin{minted}[fontsize=\small,frame=single,framesep=2mm,
                baselinestretch=1.2,fontsize=\footnotesize,
                linenos]{C++}
#include <cassert>
#include <cstdio>
#include "functions.h"

int main() {
    auto errorResult = Functions::division(1,0);
    if (errorResult.has_value()) {
        printf("result = %f\n", errorResult.value());
    } else {
        auto optionalError = errorResult.error().as<Functions::DivByZero>();
        assert(optionalError.isSome());

        auto errorValue = optionalError.get();
        assert(errorValue == Functions::DivByZero::divisorIsZero);
        errorValue.getMessage();
    }

    auto goodResult = Functions::division(4,2);
    if (goodResult.has_value()) {
        printf("result = %f\n", goodResult.value());
    } else {
        auto optionalError = goodResult.error().as<Functions::DivByZero>();
        assert(optionalError.isSome());

        auto errorValue = optionalError.get();
        errorValue.getMessage();
    }
    return 0;
}
\end{minted}
\caption{Handling Swift Error using an expected type based on the std::expected proposal.}
\label{lst:example9}
\end{listing}

The main difference in this example is that the ``try-catch'' block is not used. Instead, the function's return is tested to check if it contains a value, the good case where we just print the value or not. This last case is most interesting, as the error was returned and can be accessed by the ``error()'' function and then handled as before.

The commands below show how to compile and test this program. First, we need to compile and generate its object file with the ``-fno-exceptions'' flag, compile it with the Swift file containing the function and error definitions and then generate the executable program.

\begin{listing}[H]
\begin{minted}[fontsize=\small,frame=single,framesep=2mm,
                baselinestretch=1.2,fontsize=\footnotesize,
                linenos]{Bash}
$ clang++ -I ${SWIFT_BUILD_DIR}/swift-macosx-arm64/./lib/swift \ 
        -c Expected_Example.cpp -fno-exceptions -I /. -o expected_example.o

$ swiftc -Xfrontend -enable-experimental-cxx-interop Example.swift \
        -o expected_example -Xlinker expected_example.o -module-name Functions \
        -Xfrontend -entry-point-function-name -Xfrontend swiftMain

$ ./expected_example
\end{minted}
\caption{Commands to build, link, and execute a C++ program that calls a Swift function and can't throw exceptions, instead, it uses the Swift::Expected<T> class.}
\label{lst:exampleCommandNOExceptions}
\end{listing}


\citeoption{abnt-etal-list=5}
\bibliographystyle{abntex2-cite-min}
\bibliography{references}

%
%











\phantompart
\printindex

\end{document}